\documentstyle[12pt]{article}

\newcommand {\e} {\mbox{\rm e}}

\newcommand {\nn}    {\nonumber}
\newcommand {\vs}[1]  { \vspace*{#1 cm} }

\newcounter{eq}
\newcounter{sc}

\newcommand {\PL}   {Phys.Lett.}
\newcommand {\PR}   {Phys.Rev.}
\newcommand {\PRL}   {Phys.Rev.Lett.}

\def\overleftrightarrow#1{\vbox{\ialign{##\crcr
 $\leftrightarrow$\crcr\noalign{\kern-1pt\nointerlineskip}
 $\hfil\displaystyle{#1}\hfil$\crcr}}}

\newlength{\minitwocolumn}
\begin{document}

\begin{flushright}
EDO-EP-36\\
February, 2001\\
\end{flushright}
\vspace{30pt}

\pagestyle{empty}
\baselineskip15pt

\begin{center}
{\large\bf Locally Localized Gravity Models in Higher Dimensions

 \vskip 1mm
}

\vspace{20mm}

Ichiro Oda
          \footnote{
          E-mail address:\ ioda@edogawa-u.ac.jp
                  }
\\
\vspace{10mm}
          Edogawa University,
          474 Komaki, Nagareyama City, Chiba 270-0198, JAPAN \\

\end{center}


\vspace{15mm}
\begin{abstract}
We explore the possibility of generalizing the locally localized
gravity model in five space-time dimensions to arbitrary
higher dimensions. 
In a space-time with negative cosmological 
constant, there are essentially two kinds of higher-dimensional
cousins which not only take an analytic form but also are free 
from the naked curvature singularity in a whole bulk space-time.
One cousin is a trivial extension of five-dimensional model,
while the other one is in essence in higher dimensions.
One interesting observation is that in the latter model,
only anti-de Sitter ($AdS_p$) brane is physically meaningful 
whereas de Sitter ($dS_p$) and Minkowski ($M_p$) branes
are dismissed.  
Moreover, for $AdS_p$ brane in the latter model, we study the property 
of localization of various bulk fields on a single brane. 
In particular, it is shown that the presence of the brane cosmological 
constant enables bulk gauge field and massless fermions to confine to
the brane only by a gravitational interaction. We find a novel relation
between mass of brane gauge field and the brane cosmological
constant.

\vspace{15mm}

\end{abstract}

\newpage
\pagestyle{plain}
\pagenumbering{arabic}


\rm
\section{Introduction}

In recent years, an idea that our world is a 3-brane embedded in a 
higher dimensional space-time, with some of extra dimensions being 
macroscopically large, has attracted a lot of attention as a resolution
of the hierarchy problem, the supersymmetry breaking and the cosmological
constant problem and so on. 
In particular, Randall and Sundrum have found a solution to the five 
dimensional Einstein's equations with a Minkowski flat 3-brane in $AdS_5$ 
and shown that the effects of the four dimensional gravity on the brane 
are reproduced without the need to compactify the fifth 
dimension \cite{Randall1, Randall2}.
(This model was generalized to the case of many branes in 
Ref. \cite{Oda1, Oda2}.)  

One disadvantage in Randall-Sundrum model \cite{Randall1} is the presence 
of a brane with negative tension. Although this brane is located at a fixed
point of $S^1/Z_2$ orbifold in such a way that the fluctuation modes
associated with the brane, which are necessarily physical ghost modes,
do not appear, the existence of the negative tension brane violates
the weak energy theorem in the bulk \cite{Witten}.

Another disadvantage in the Randall-Sundrum model is related to the
localization
of bulk fields on a brane \cite{Oda3}. 
In the conventional brane world scenario, the Standard Model gauge and matter 
fields are assumed to be localized on our brane, 
whereas gravity freely propagates in a bulk space-time. But this assumption
is quite unnatural since we tacitly discriminate gravity from the other
fields.
Since the graviton corresponds to the fluctuation mode of the space-time 
geometry, it automatically sees the whole structure of the space-time and
consequently lives in the bulk space-time.
The physically plausible setup is then to treat the Standard Model gauge and 
matter fields on an equal footing with gravity and consider all the local
fields as
the fields living in the bulk space-time. From this context, we can regard 
the Randall-Sundrum model \cite{Randall2} as a successful model for the 
localization of gravity on a brane. However, it is well known that in the
original Randall-Sundrum model it is very difficult to localize the gauge
fields \cite{Dvali, Pomarol, Rizzo, Bajc} and the massless fermions, those
are, 
spin-1/2 massless Dirac spinor \cite{Jackiw, Grossman, Chang, Randjbar} and
spin-3/2 massless gravitino \cite{Oda4}, on a brane by a gravitational 
interaction.

Recently, there has been an interesting development which circumvents
simultaneously
the two disadvantages mentioned above \cite{Kogan, Karch, Miemiec, Schwartz, 
Oda5, Tachibana}. The models deal with a single or two positive tension 
anti-de Sitter $AdS_4$ brane(s) in a five dimensional anti-de Sitter
space-time
$AdS_5$, where four dimensional gravity is induced on the $AdS_4$ brane 
owing to the localization of a massive and normalizable bound state
\cite{Karch}.
Moreover, it was shown that all the Standard Model particles 
are localized on the $AdS_4$ brane only through the gravitational interaction
\cite{Oda5}. For instance, the appearance of
zero-mode with dependence of a fifth dimension supplies us with a novel
mechanism
for the localization of the bulk gauge field on the brane.
Even if nature seems to favor a Minkowski brane $M_4$ with zero cosmological 
constant rather than an $AdS_4$ with negative cosmological 
constant, it is impossible to rule out the possibility
that our world might have a very tiny negative cosmological constant which is
consistent with the present observations. Interestingly enough, in the
$AdS_4$ brane 
model the existence of a $\it{massless}$ 'photon' on a brane demands that the 
brane cosmological constant must be small in Planck units enough not to
violate 
experiment \cite{Oda5}.

The aim of this paper is to generalize this interesting model to higher
dimensions.
Such a generalization is of course of importance from the viewpoint of a
underlying
fundamental theory in higher dimensions such as ten dimensional superstring
theory. 

We will regard
the branes as $\it{global}$ defects with the number $p$ of longitudinal
dimensions
in a higher dimensional space-time with $D$ bulk dimensions and $n$ 
extra transverse ones (so the equality $D = p + n$ holds). A set of $n$
scalar fields 
with the Higgs potential, thereby breaking the $\it{global}$ $SO(n)$
symmetry to 
$SO(n-1)$ symmetry, are utilized to generate the $\it{global}$ defects 
\cite{Vilenkin, Olasagasti}. A topological argument $\Pi_{n-1}(SO(n-1)) = Z$,
which expresses the fact that a mapping of configuration space at spatial
infinity
to a vacuum manifold is topologically nontrivial, guarantees the stability
of the
defects under deformations. 
 
The plan of the paper is as follows. In the next section we review the
model setup
and then in section 3 we look for solutions to Einstein's equations. 
In section 4, we study the metric 
fluctuations and derive a Schrodinger-like equation. Because of a complicated
form of the solution and the lack
of the knowledge inside the core, it is difficult to understand an exact
formula
of Newton's potential so we shall be contented with some qualitative
understanding of the
solution. 
In section 5, we show that the zero-mode of bulk gauge field is normalizable
owing to the presence of the cosmological constant, thereby leading to the
localization
of gauge field on a brane. But it is shown that the localization is not so
sharp on
the brane and spreads rather widely in a bulk. 
Section 6 is devoted to the treatment of fermionic fields.
Discussions and future works are summarized in section 7.

\section{Model setup}

In this section, we shall review the construction of '$\it{global}$'
topological
defect model in higher dimensions \cite{Vilenkin, Olasagasti}.
The solutions to Einstein's equations which we shall derive below
can be found in essence in the article of Olasagasti and Vilenkin
\cite{Vilenkin, Olasagasti}, but we shall not only derive the solutions
in a more unified metric ansatz but also examine their physical properties
in detail from a different viewpoint. 
In this paper, we shall follow the notations and the conventions 
in our previous papers \cite{Oda6}.

The action with which we start is that of gravity
in general $D$ dimensions, with the conventional Einstein-Hilbert
action and some matter action which will be specified later:
\begin{eqnarray}
S = \frac{1}{2 \kappa_D^2} \int d^D x  
\sqrt{-g} \left(R - 2 \Lambda \right) 
+ \int d^D x  \sqrt{-g} L_m.
\label{1}
\end{eqnarray}
Taking the variation of the action (\ref{1}) with respect to the
$D$-dimensional
metric tensor $g_{MN}$ we obtain Einstein's equations in $D$ dimensions:
\begin{eqnarray}
R_{MN} - \frac{1}{2} g_{MN} R 
= - \Lambda g_{MN}  + \kappa_D^2 T_{MN},
\label{2}
\end{eqnarray}
where the energy-momentum tensor is defined as
\begin{eqnarray}
T_{MN} = - \frac{2}{\sqrt{-g}} \frac{\delta}{\delta g^{MN}}
\int d^D x \sqrt{-g} L_m.
\label{3}
\end{eqnarray}

To find the spherically symmetric solutions in the bulk, we shall adopt 
the following metric ansatz:
\begin{eqnarray}
ds^2 &=& g_{MN} dx^M dx^N  \nn\\
&=& g_{\mu\nu} dx^\mu dx^\nu + dr^2 + g_{mn} dy^m dy^n  \nn\\
&=& e^{-A(r)} \hat{g}_{\mu\nu} dx^\mu dx^\nu + dr^2 
+ e^{-B(r)} d \Omega_{n-1}^2,
\label{4}
\end{eqnarray}
where $M, N, ...$ denote $D$-dimensional space-time indices, 
$\mu, \nu, ...$ $p$-dimensional brane ones, and $m, n, ...$
$(n-1)$-dimensional extra spatial ones, so the equality $D=p+n$
holds. (We assume $p \ge 4$.) We sometimes denote $g_{mn} = \e^{-B(r)}
\tilde{g}_{mn}(x^l)$. Note that the reason why we take account of 
this metric ansatz comes from the holographic principle where the
'radial' coordinate $r$ plays the role of scale of the $AdS$ renormalization
group, so it is straightforward to extend various results of AdS/CFT
correspondence such as "c-theorem" to the present case. 
Moreover, we shall take an ansatz for the energy-momentum tensor
respecting the spherical symmetry:
\begin{eqnarray}
T^\mu_\nu &=& \delta^\mu_\nu t_o(r),  \nn\\
T^r_r &=& t_r(r),  \nn\\
T^{\theta_2}_{\theta_2} &=& T^{\theta_3}_{\theta_3} = \cdots
= T^{\theta_n}_{\theta_n} = t_\theta(r),
\label{5}
\end{eqnarray}
where $t_i(i=o, r, \theta)$ are functions of only the radial
coordinate $r$. 

Under these ansatzs, after a straightforward calculation,
Einstein's equations reduce to the forms
\begin{eqnarray}
e^A \hat{R} - \frac{p(n-1)}{2} A' B' - \frac{p(p-1)}{4} (A')^2
- \frac{(n-1)(n-2)}{4} (B')^2 \nn\\
+ (n-1)(n-2) e^B - 2\Lambda + 2 \kappa_D^2 t_r = 0,
\label{6}
\end{eqnarray}
\begin{eqnarray}
e^A \hat{R} + (n-2) B'' - \frac{p(n-2)}{2} A' B' 
- \frac{(n-1)(n-2)}{4} (B')^2  \nn\\
+ (n-2)(n-3) e^B + p A'' -  \frac{p(p+1)}{4} (A')^2 - 2\Lambda 
+ 2 \kappa_D^2 t_\theta = 0,
\label{7}
\end{eqnarray}
\begin{eqnarray}
\frac{p-2}{p} e^A \hat{R} + (p-1)(A'' - \frac{n-1}{2} A' B')
- \frac{p(p-1)}{4} (A')^2 \nn\\
+ (n-1) [B'' - \frac{n}{4} (B')^2 + (n-2) e^B ]  
- 2\Lambda + 2 \kappa_D^2 t_o = 0,
\label{8}
\end{eqnarray}
where the prime denotes the differentiation with respect to $r$,
and $\hat{R}$ is the scalar curvature associated with the
brane metric $\hat{g}_{\mu\nu}$.
Here we define the cosmological constant on the $(p-1)$-brane, 
$\Lambda_p$, by the equation
\begin{eqnarray}
\hat{R}_{\mu\nu} - \frac{1}{2} \hat{g}_{\mu\nu} \hat{R} 
= - \Lambda_p \hat{g}_{\mu\nu}.
\label{9}
\end{eqnarray}
In addition, the conservation law for the energy-momentum tensor,
$\nabla^M T_{MN} = 0$ takes the form
\begin{eqnarray}
t'_r = \frac{p}{2} A' (t_r - t_o) + \frac{n-1}{2} B' (t_r - t_\theta).
\label{10}
\end{eqnarray}

The formulation reviewed thus far \cite{Oda6} is rather general 
in that we have assumed only the metric ansatz (\ref{4}). 
Here let us specify the model by fixing the matter action.
Following Refs. \cite{Vilenkin, Olasagasti}, we shall take a
multiplet of $n$ scalar fields $\Phi^a$ with the Higgs potential:
\begin{eqnarray}
L_m =  - \frac{1}{2} g^{MN} \partial_M \Phi^a \partial_N \Phi^a 
+ \frac{\lambda}{4}(\Phi^a\Phi^a - \eta^2)^2 ,
\label{11}
\end{eqnarray}
from which the energy-momentum tensor takes the form
\begin{eqnarray}
T_{MN} = \partial_M \Phi^a \partial_N \Phi^a  - \frac{1}{2} \ g_{MN} 
\partial_P \Phi^a \partial^P \Phi^a + \ g_{MN} \frac{\lambda}{4}
(\Phi^a\Phi^a - \eta^2)^2.
\label{12}
\end{eqnarray}

Then, the familiar 'hedgehog' ansatz leads to a global defect
\begin{eqnarray}
\Phi^a =  f(r) \hat{r}^a,
\label{13}
\end{eqnarray}
where $\hat{r}^a$ is the unit vector on the $(n-1)$-sphere and the function
$f(r)$ takes the form
\begin{eqnarray}
f(0) = 0, \ \lim_{r \rightarrow \infty} f(r) = \eta.
\label{14}
\end{eqnarray}
Namely, it is considered that the defect has $\Phi^a = 0$ at the center
of the core and approaches the radial 'hedgehog' configuration
$\Phi^a =  \eta \hat{r}^a$ outside the core. In this paper, we limit
ourselves to the exterior solutions, where the configuration is
given by $\Phi^a =  \eta \hat{r}^a$. Note that this configuration
becomes an accurate approximation as the coupling constant $\lambda$
gets large. A big question about 'global' defects is whether there could
be a stable localized core or not \footnote{In the absence of gravity, in
other
words, in the Minkowski space, Virial theorem tells us that for 
$D \ge 3$ there are no such static solutions. This theorem is circumvented
when gravity switches on and there is a negative cosmological constant
as in the case at hand.}.
This problem is closely connected with
physics inside the core so now we cannot answer this important problem.

\section{Solutions}

In this section, we solve a set of Einstein's equations (\ref{6})-(\ref{8})
derived in the previous section. 
In this paper, we pay attention to only the case 
of the bulk cosmological constant being negative, $\Lambda < 0$ 
in order to search higher dimensional analogs corresponding to 
an $AdS_4$ brane solution in $AdS_5$ \cite{Karch}.

First of all, let us notice that with the ansatz 
$\Phi^a =  \eta \hat{r}^a$ which holds only outside the defect core, 
the energy-momentum tensor takes the forms 
\begin{eqnarray}
t_o = t_r = -\frac{1}{2} (n-1) \eta^2 e^{B(r)}, \
t_\theta = -\frac{1}{2} (n-3) \eta^2 e^{B(r)},
\label{15}
\end{eqnarray}
which obviously satisfy the conservation law (\ref{10}).

Next, to find analytic solutions we need to set up a more
specific metric ansatz, for which we shall take the form
\begin{eqnarray}
B(r) = c A(r) + d, \ R_0^2 \equiv \e^{-d},
\label{16}
\end{eqnarray}
where $c$ and $d$ (or $R_0$) are constants, which will be later fixed
by Einstein's equations.
Then it is straightforward to solve Einstein's equations 
(\ref{6})-(\ref{8}) whose solutions can be divided into two
kinds of cousins. One cousin, being a trivial extension of branes in 
$AdS_5$, belongs to a class having $c = 0$. 
According to the signature of
the brane cosmological constant, let us separate this class of solutions 
to three branes' solutions, de-Sitter brane $dS_p$, Minkowski brane
$M_p$, and anti de-Sitter brane $AdS_p$ \footnote{In this paper, 
we consider only the maximally symmetric solutions on a brane.}.

(i) $dS_p$ brane:
\begin{eqnarray}
ds^2 &=& \sinh^2 \omega r d\hat{s}_{+}^2 + dr^2 + R_0^2 d \Omega_{n-1}^2
\nn\\
\hat{R} &=& - 2 \Lambda \frac{p-1}{n+p-2}>0, \ 
\Lambda_{dS} = - \Lambda \frac{(p-1)(p-2)}{p(n+p-2)}>0.
\label{17}
\end{eqnarray}

(ii) $M_p$ brane:
\begin{eqnarray}
ds^2 &=& \e^{\mp 2 \omega r} d\hat{s}_{0}^2 + dr^2 + R_0^2 d \Omega_{n-1}^2
\nn\\
\hat{R} &=& \Lambda_M = 0.
\label{18}
\end{eqnarray}

(iii) $AdS_p$ brane:
\begin{eqnarray}
ds^2 &=& \cosh^2 \omega r d\hat{s}_{-}^2 + dr^2 + R_0^2 d \Omega_{n-1}^2
\nn\\
\hat{R} &=& 2 \Lambda \frac{p-1}{n+p-2}<0, \ 
\Lambda_{AdS} = \Lambda \frac{(p-1)(p-2)}{p(n+p-2)}<0.
\label{19}
\end{eqnarray}
Here $\omega, R_0^2$ are respectively given by
\begin{eqnarray}
\omega = \sqrt{\frac{-2 \Lambda}{p(n+p-2)}}, \ 
R_0^2 = \frac{1}{2 \Lambda} (n+p-2)(n-2-\kappa_D^2 \eta^2),
\label{20}
\end{eqnarray}
where $R_0^2 > 0$ requires $n-2-\kappa_D^2 \eta^2 < 0$. This class
of solutions have been first derived in Ref. \cite{Vilenkin}. The 
common feature in this class is that the $(n-1)$-sphere has 
a constant radius $R_0$, because of which we have called it 
a $\it{trivial}$ extension of branes in $AdS_5$ in the above. 
Indeed, it is easy to show that this class of solutions share the
same properties such as the corrections to Newton's law as for 
corresponding five dimensional cases. 
Thus we shall not consider this class of solutions any more in this
paper.

A different class of solutions are provided when $c=1$. Again 
we shall present the solutions below by following the signature of
the brane cosmological constant.

(i) $dS_p$ brane:
\begin{eqnarray}
ds^2 &=& \sinh^2 \omega r d\hat{s}_{+}^2 + dr^2 + R_0^2 
\sinh^2 \omega r d \Omega_{n-1}^2
\nn\\
\hat{R} &=& - 2 \Lambda \frac{p}{n+p-1}>0, \ 
\Lambda_{dS} = - \Lambda \frac{p-2}{n+p-1}>0, \
n-2-\kappa_D^2 \eta^2 > 0.
\label{21}
\end{eqnarray}

(ii) $M_p$ brane:
\begin{eqnarray}
ds^2 &=& \e^{\mp 2 \omega r} d\hat{s}_{0}^2 + dr^2 + R_0^2 
\e^{\mp 2 \omega r} d \Omega_{n-1}^2
\nn\\
\hat{R} &=& \Lambda_M = 0, \ 
n-2-\kappa_D^2 \eta^2 = 0.
\label{22}
\end{eqnarray}

(iii) $AdS_p$ brane:
\begin{eqnarray}
ds^2 &=& \cosh^2 \omega r d\hat{s}_{-}^2 + dr^2 + R_0^2 
\cosh^2 \omega r d \Omega_{n-1}^2
\nn\\
\hat{R} &=&  2 \Lambda \frac{p}{n+p-1}<0, \ 
\Lambda_{AdS} = \Lambda \frac{p-2}{n+p-1}<0, \
n-2-\kappa_D^2 \eta^2 < 0.
\label{23}
\end{eqnarray}
Here $\omega, R_0^2$ are respectively given by
\begin{eqnarray}
\omega = \sqrt{\frac{-2 \Lambda}{(n+p-2)(n+p-1)}}, \ 
R_0^2 = - \frac{1}{2 \Lambda} (n+p-2)|n-2-\kappa_D^2 \eta^2|,
\label{24}
\end{eqnarray}
but in the case of the Minkowski brane $M_p$, $R_0$ is a free parameter.
This class of solutions have been also in essence derived in 
Ref. \cite{Vilenkin} but with a different metric ansatz from ours.
Note that one advantage of our metric ansatz (\ref{16}) over the ones
in Ref. \cite{Vilenkin} is that we have derived two classes of solutions
in a unified way, while the authors in Ref. (\ref{16}) 
have set up different metric ansatzs
and needed the change of variables to reach the forms listed in
the above.

Now let us attempt to understand the solutions (\ref{21})-(\ref{23}) in 
more detail. To do so, let us calculate the 
D-dimensional scalar curvature under the ansatz (\ref{4}) whose
result is given by
\begin{eqnarray}
R &=& g^{MN} R_{MN} \nn\\
&=& \e^A \hat{R} + p A'' + (n-1) B'' - \frac{p(p+2)}{4}(A')^2
- \frac{p(n-1)}{2} A' B' - \frac{n(n-1)}{4}(B')^2 \nn\\
&+& (n-1)(n-2) \e^B.
\label{25}
\end{eqnarray}
In particular, the last term in $R$ reveals that the cases of 
$n = 1, 2$ are qualitatively different from higher dimensional cases 
$n \ge 3$. The reason is that for $n=1$ (domain wall) 
the extra space is flat and for $n=2$ (string-like defect)
the extra space is still conformally flat, while for
$n \ge 3$ the extra space is essentially curved \cite{Gherghetta}.
The presence of this term makes many solutions to Einstein's
equations in higher dimensions physically uninteresting owing to
the appearance of the naked curvature singularity in the bulk space-time. 
Some people do not regard the appearance of the naked curvature singularity
as a sick property of solutions by taking the optimistic attitude that such a 
singularity would be smoothed by quantum effects or string theory corrections.
In contrast, we consider the naked curvature singulariy to be a serious 
problem of solutions and take a strict criterion such that 
classical solutions to Einstein's equations should be free from the naked 
curvature singularity \footnote{The existence
of the curvature singularity at the origin $r=0$ might be admissible since 
in some cases this singularity could be identified with core of 
the brane \cite{Charmousis}.}. 

Imposing the singularity-free condition as the physical requirement, for $n
\ge 3$
$dS_p$ brane in Eq. (\ref{21}) must be dismissed from physical solutions. 
Note that in this case, the real problem is that the line element is singular 
at $r=0$ $\it{even}$ in the absence of a defect ($\eta = 0$) \cite{Vilenkin}.
{}For $n=2$, $dS_p$ brane is not the solution owing to the relation 
$n-2-\kappa_D^2 \eta^2 > 0$ when there is no defect ($\eta = 0$), so we
also dismiss this case. Of course, for $n=1$, $dS_p$ brane is physical and
corresponds to the $dS$ domain wall solution.

Next, in $M_p$ brane Eq. (\ref{22}), for $n \ge 3$, the solution with the
upper
sign has the naked curvature singulaity at the spatial infinity, so we dismiss
this solution. On the other hand, the solution with the lower sign is free
from
the curvature singularity, but it turns out that the solution cannot 
localize gravity on a defect,
so we also dismiss this case. The remaining possibilities are when $n = 1, 2$.
{}For $n=2$, the solution corresponds to Gregory's solution \cite{Gregory,
Gherghetta2, Oda6} and as seen from the relation $n-2-\kappa_D^2 \eta^2 = 0$
this solution describes a $\it{local}$ string-like defect so we also dismiss
this solution from our present consideration. The solution in the case of 
$n=1$ is nothing but the Randall-Sundrum solution \cite{Randall1, Randall2}
(when $p=4$).

We are ready to analyse $AdS_p$ brane Eq. (\ref{23}) in a similar manner.
{}For $n=1$, the solution obviously corresponds to an $AdS_p$ brane in
$AdS_{p+1}$
\cite{Karch}. Note that for $n \ge 2$ the solution is completely free from
the curvature singularity and constitutes of a higher dimensional, 
nontrivial extension of an $AdS_p$ brane in $AdS_{p+1}$. 
Remarkably, as shown later, this solution 
localizes all local bulk fields on a defect only through the gravitational
interaction.

Before closing this section, let us summarize the results obtained in this
section.
We have derived two classes of classical solutions to Einstein's equations
in higher dimensions.
One class of solutions are a trivial extension of the domain wall solutions.
The other class of solutions are a nontrivial extension, but almost all
solutions except $AdS_p$ brane are unphysical because of the existence of
the naked curvature singularity and the non-localization of gravity on a
defect. It is rather surprising that in higher dimensions ($n \ge 2$) only 
the $AdS_p$ brane solution is selected as physical solution, while in
the case of $n=1$ domain wall three types of brane, those are, $dS_p$, $M_p$
and $AdS_p$ branes are permissible.

\section{Gravitational fluctuations}

In the following sections, we shall turn our attention to the properties
of an $AdS_p$ brane solution (\ref{23}) in higher dimensional space-time.
The aim of this section is to study the gravitational fluctuations around
the background (\ref{23}).

First, let us rewrite the metric (\ref{23}) in terms of the conformal 
coordinates
\begin{eqnarray}
ds^2 &=& \cosh^2 \omega r \hat{g}_{\mu\nu} dx^\mu dx^\nu + dr^2 + R_0^2 
\cosh^2 \omega r d \Omega_{n-1}^2
\nn\\
&=&  \e^{-A(z)}  \left( \hat{g}_{\mu\nu} dx^\mu dx^\nu + dz^2 + R_0^2 
d \Omega_{n-1}^2 \right),
\label{26}
\end{eqnarray}
where $\e^{-A(z)}$ and the relation between two coordinate symtems
are respectively given by
\begin{eqnarray}
\e^{-A(z)} &=& \frac{1}{\sin^2 \omega z}, \nn\\
\e^{\omega r} &=& \tan \frac{1}{2} \omega z.
\label{27}
\end{eqnarray}
Since the 'radial' coordinate $r$ runs from $0$ to $\infty$, this relation 
yields the range of $z$, which is $\frac{\pi}{2 \omega} \le z \le
\frac{\pi}{\omega}$. 

We will only consider the transverse, traceless fluctuations around the
background metric (\ref{26}) in the conformal $z$-coordinates
\begin{eqnarray}
ds^2 = \e^{-A(z)}  \left[ (\hat{g}_{\mu\nu} + h_{\mu\nu}(x^M)) dx^\mu dx^\nu 
+ dz^2 + R_0^2 d \Omega_{n-1}^2 \right],
\label{28}
\end{eqnarray}
where $\nabla^\mu h_{\mu\nu} = g^{\mu\nu} h_{\mu\nu} = 0$. Then, it is 
straightforward to show that Einstein's equations reduce to the form
of the linearized equations
\begin{eqnarray}
\frac{1}{\sqrt{-g}} \partial_M (\sqrt{-g} g^{M N} \partial_N h_{\mu\nu})
- 2 \Lambda h_{\mu\nu} = 0.
\label{29}
\end{eqnarray}
Given the symmetries of the background metric, we separate variables
as
\begin{eqnarray}
h_{\mu\nu}(x^M) = \phi_{\mu\nu}(x^\mu) \check{Z}_{lm}(z) Y_{lm_i}
(\Omega),
\label{30}
\end{eqnarray}
where $Y_{lm_i}(\Omega)$ are the spherical harmonics for the 
$(n-1)$-sphere with eigenvalue $\Delta_l = l(l+n-2)$.  
And $\phi_{\mu\nu}(x^\mu)$ satisfy the equations of motion
$(\hat{\Box} - 2 \Lambda \e^{-A})\phi_{\mu\nu} = m_0^2 \phi_{\mu\nu}$
with the definition of $\hat{\Box} \equiv \frac{1}{\sqrt{-\hat{g}}} 
\partial_\mu (\sqrt{-\hat{g}} \hat{g}^{\mu\nu} \partial_\nu)$.
The equations (\ref{29}) then reduce to
\begin{eqnarray}
\e^{\frac{D-2}{2}A} \partial_z (\e^{\frac{D-2}{2}A} \partial_z
\check{Z}_{lm}) + m^2 \check{Z}_{lm} = 0,
\label{31}
\end{eqnarray}
where $m^2 = m_0^2 + \frac{\Delta_l}{R_0^2}$. After changing to a
new function 
\begin{eqnarray}
\check{Z}_{lm} = \e^{\frac{D-2}{4}A} Z_{lm},
\label{32}
\end{eqnarray}
we find a Schrodinger-like equation for $Z_{lm}$:
\begin{eqnarray}
[-\partial_z^2 + V(z)] Z_{lm}(z) = m^2 Z_{lm}(z),
\label{33}
\end{eqnarray}
where the potential is of the form
\begin{eqnarray}
V(z) = \frac{(D-2)^2}{16} (A')^2 - \frac{D-2}{4} A'',
\label{34}
\end{eqnarray}
with the prime denoting the differentiation with respect to $z$.
If we introduce a new variable $w \equiv \omega z$, taking
the range $\frac{\pi}{2} \le w \le \pi$, instead of $z$ and make
use of the concrete expression of $A(z)$ in (\ref{27}), we
finally arrive at the equation:
\begin{eqnarray}
[-\partial_w^2 + U(w)] Z(w) = E Z(w),
\label{35}
\end{eqnarray}
where we have omitted to write the indices $l, m$ on $Z(w)$ explicitly, and
$U(w)$ and $E$ are respectively defined as
\begin{eqnarray}
U(w) &=& -\frac{(D-2)^2}{4} + \frac{D(D-2)}{4} \frac{1}{\sin^2 w},
\nn\\
E &=& \frac{m^2}{\omega^2}.
\label{36}
\end{eqnarray}

To have the second-rank linear differential equation well-defined,
we need to impose boundary conditions at $w = \frac{\pi}{2}, \pi$.
The boundary condition at $w = \pi$ is the Dirichlet condition,
$Z(\pi)=0$ since the potential $U(w)$ becomes an infinity there.
The delicate problem is what boundary condition we have to impose
at $w = \frac{\pi}{2}$ where there is the core of a topological 
defect. We have only solved Einstein's equations in the exterior
region outside the core so that in principle we have no knowledge
about physics inside the core, which makes it difficult to set up
the boundary condition at $w = \frac{\pi}{2}$. 
However, the condition that the
differential operator should be self-adjoint, which is necessary
for $Z$ to have a complete basis, requires us to choose a homogeneous
boundary condition at $w = \frac{\pi}{2}$
\begin{eqnarray}
\xi_1 Z'(\frac{\pi}{2}) + \xi_2 Z(\frac{\pi}{2}) = 0,
\label{37}
\end{eqnarray}
where $\xi_1, \xi_2$ are constants and the prime now denotes the 
differentiation with respect to $w$. Thus physics inside the core of a 
defect should satisfy this boundary condition to have a smooth continuity
between inside and outside the core of a defect. (Here for simplicity we
have neglected the core size.)

With these results in hand, we are ready to study a solution to
the equation (\ref{35}). After some elementary manipulation 
\cite{Kogan, Karch, Miemiec, Schwartz}, a solution is given
by a linear combination of Gauss's hypergeometric function $F$:
\begin{eqnarray}
Z &=& \frac{A_1}{(\sin w)^{\frac{D-2}{2}}} F(-\frac{D-2}{4} 
+ \frac{\sqrt{4E+(D-2)^2}}{4}, -\frac{D-2}{4} - \frac{\sqrt{4E+(D-2)^2}}{4},
\frac{1}{2}; \cos^2 w)    \nn\\
&+& \frac{A_2 \cos w}{(\sin w)^{\frac{D-2}{2}}} F(-\frac{D-4}{4} 
+ \frac{\sqrt{4E+(D-2)^2}}{4}, -\frac{D-4}{4} - \frac{\sqrt{4E+(D-2)^2}}{4},
\frac{3}{2}; \cos^2 w), \nn\\
\label{38}
\end{eqnarray}
where $A_1, A_2$ are integration constants.
At this stage, given the boundary condition $Z(\pi)=0$, we find
an equation
\begin{eqnarray}
A_2 =  2 A_1 \frac{\Gamma(\frac{D+2}{4} - \frac{\sqrt{4E+(D-2)^2}}{4})
\Gamma(\frac{D+2}{4} + \frac{\sqrt{4E+(D-2)^2}}{4})}
{\Gamma(\frac{D}{4} - \frac{\sqrt{4E+(D-2)^2}}{4})
\Gamma(\frac{D}{4} + \frac{\sqrt{4E+(D-2)^2}}{4})}.
\label{39}
\end{eqnarray}

The remaining work is to impose the boundary condition (\ref{37}) to
fix the integration constants $A_1, A_2$, but it is a very delicate
problem because of a complicated structure of Gauss's hypergeometric 
function and in consequence only the numerical analysis is available.
We leave this numerical analysis for a future work.
Here we shall investigate only a set of massive excited states
which are supported by $U(w)$. Then the natural choice of the
boundary condition at $w = \frac{\pi}{2}$ is the Neumann boundary condition 
$Z'(\frac{\pi}{2}) = 0$. Together with this boundary condition and
Eq. (\ref{39}), we obtain $A_2 = 0$ and eigenvalues of Eq. (\ref{35})
\begin{eqnarray}
E_k = k (k + D - 2),
\label{40}
\end{eqnarray}
where $k = 1, 2, \cdots$. This equation then gives the natural
higher dimensional generalization of mass formula of KK states
in $AdS_4$ brane \cite{Karch}
\begin{eqnarray}
m_k^2 = - \frac{2}{(p-2)(n+p-2)} k (k + D - 2) \Lambda_{AdS},
\label{40A}
\end{eqnarray}
where we have used (\ref{23}), (\ref{24}) and (\ref{36}).
Hence, in the $\Lambda_{AdS} \rightarrow 0$ limit, these states
become massless degenerate states, thereby giving rise to corrections
to Newton's law whose size is of the order 
$\cal{O}$($\sqrt{\Lambda_{AdS}}$). 

On the other hand, a massive bound state, which is trapped and 
generates gravity on an $AdS$ brane, is supported by the attractive 
potential around the core. In the model at hand, 
the information about this
attractive potential is implicitly included in the boundary condition
(\ref{37}). Recall that the boundary condition of $AdS_4$ brane in
$AdS_5$ certainly satisfies this equation. Anyway in order to understand
this problem completely, it would be necessary to construct a 
physically plausible core model.

\section{Localization of gauge fields}

In this section we are willing to consider the localization of spin-1 
$U(1)$ vector field on an $AdS_p$ brane (\ref{23}). Incidentally
the generalization to the nonabelian gauge fields is straightforward.
As mentioned in section 1, it is well known that in the original
Randall-Sundrum model the gauge fields cannot be localized on a
domain wall by the gravitational interaction \cite{Dvali, Pomarol, Rizzo, 
Bajc}. Since we have various gauge fields in our world, the impossibility
of confining gauge fields to a brane imposes a serious drawback on
the scenario of brane world. Of course, there might be some ingeneous
mechanism for the localization of gauge fields by invoking additional
interactions except the gravitational one \cite{Dvali}, 
but we believe that such
a mechanism is artificial and the universal interaction, that is, the
gravitational interaction should provide us with the localization 
mechanism for the whole local fields including gauge fields. 
From this context, it is of interest
to ask whether or not $AdS_p$ brane (\ref{23}) presents the localization
mechanism for the gauge fields.

Let us start with the familiar action of $U(1)$ gauge field
\begin{eqnarray}
S_1 = -\frac{1}{4} \int d^D x \sqrt{-g} g^{MN} g^{RS} F_{MR} F_{NS}, 
\label{41}
\end{eqnarray}
where $F_{MN} = \partial_M A_N - \partial_N A_M$. The equations of motion
become $\partial_M (\sqrt{-g} g^{MN} g^{RS} F_{NS}) = 0$. 
To study the localization of the gauge field, it is convenient to
use the conformal $z$-coordinates (\ref{26}). In the coordinates,
for simplicity, we shall focus on only the brane gauge field $A_\mu(x^M)$
and set $A_z = A_{\theta_i} = 0$. Then, we look for a solution with the
form of $A_\mu(x^M) = a_\mu(x^\lambda) u(z) \chi(y^m)$ where $y^m$ denote
the angular coordinates. Here we assume the following equations of 
motion $\hat{\nabla}^\mu a_\mu = \partial^\mu f_{\mu\nu} = 
\partial_m (\sqrt{\tilde{g}} \tilde{g}^{mn} \partial_n \chi) = 0$
where $f_{\mu\nu} = \partial_\mu a_\nu - \partial_\nu a_\mu$.
With these ansatzs, the equations of motion reduce to a single 
differential equation:
\begin{eqnarray}
\partial_z \left[e^{(- \frac{D}{2} + 2) A(z)} \partial_z u(z)
\right] = 0. 
\label{42}
\end{eqnarray}

In the case of a Minkowski brane, we have selected a constant
zero-mode solution $u(z) = const$, which leads to non-localization
of the vector fields \cite{Bajc}. On the other hand, in the case 
of an $AdS$ brane, a new solution is available, which is given by 
$e^{(- \frac{D}{2} + 2) A(z)} \partial_z u(z) = const \not= 0$.
(Note that this solution is not localized on a Minkowski brane,
either.)  As a result, we obtain the following solution to Eq. (\ref{42}).
{}For $D = 2k + 3$ ($k = 1, 2, 3, \cdots$),
\begin{eqnarray}
u(z)  = \frac{\alpha}{\omega} \frac{(-1)^k}{2^{2(k-1)}} 
\sum_{l=0}^{k-1} (-1)^l {{2k-1} \choose l} \frac{\cos[(2k-2l-1)\omega z]}
{2k-2l-1} + \beta, 
\label{43}
\end{eqnarray}
and for $D = 2k + 4$ ($k = 1, 2, 3, \cdots$), 
\begin{eqnarray}
u(z)  = \frac{\alpha}{\omega} \frac{(-1)^k}{2^{2k}} 
\left\{ \sum_{l=0}^{k-1} (-1)^l {{2k} \choose l} \frac{\sin[2(k-l)\omega z]}
{k-l} + (-1)^k {{2k} \choose k} \omega z \right\} + \beta, 
\label{44}
\end{eqnarray}
where $\alpha, \beta$ are integration constants.

We would like to investigate whether this solution is localized on an $AdS_p$
brane or not. 
The substitution of this solution into the action leads to
\begin{eqnarray}
S_1^{(0)} &=& -\frac{1}{4} \int d^D x \sqrt{-g} g^{MN} g^{RS} F_{MR}^{(0)}
F_{NS}^{(0)} \nn\\ 
&=& -\frac{1}{4} \int d^px dz d^{n-1}y \sqrt{-\hat{g}} \sqrt{\tilde{g}}
\e^{(-\frac{D}{2}+2)A(z)} \times \nn\\
&{}& \left[
u^2 \chi^2 \hat{g}^{\mu\nu} \hat{g}^{\rho\sigma} f_{\mu\rho}f_{\nu\sigma}
+ 2(\partial_z u)^2 \chi^2 \hat{g}^{\mu\nu} a_\mu a_\nu
+ 2 u^2 \tilde{g}^{mn} \partial_m \chi \partial_n \chi 
\hat{g}^{\mu\nu} a_\mu a_\nu
\right]
\label{45}
\end{eqnarray}
Here we have carefully kept the KK-mass term since we wish to examine later
whether this solution leads to $\it{massless}$ 'photon' on a brane. 
The localization condition of this mode on a brane requires
the integral over $z$ in front of the kinetic term to be finite since 
the integral over the angular variables $\int d^{n-1}y \sqrt{\tilde{g}}
\chi(y)^2$ is in general finite. 
Thus let us consider this integral first: 
\begin{eqnarray}
I_1 &=& \int dz 
e^{(- \frac{D}{2} + 2)A(z)} u^2(z) \nn\\
&=& \int_{\frac{\pi}{2 \omega}}^{\frac{\pi}{\omega}} dz
\frac{1}{(\sin \omega z)^{D-4}} u^2(z).
\label{46}
\end{eqnarray}
The expressions (\ref{43}), (\ref{44}) for $u(z)$ become more complicated 
as the number of space-time dimensions gets larger, so below
we shall present explicitly only the results of the two simplest cases
$D=5, 6$, 
belonging to each branch of solutions 
although we have exmained some remaining lower dimensional cases and found 
similar results and repeated pattern depending on $D=2k+3$ or $D=2k+4$.

In the case of $D=5, p=4, n=1$, that is, an $AdS_4$ brane in $AdS_5$
\cite{Karch}, 
from Eq. (\ref{46}) the integral $I_1^{D=5}$ reads
\begin{eqnarray}
I_1^{D=5} = \int_{\frac{\pi}{2 \omega}}^{\frac{\pi}{\omega}} dz
\frac{1}{\sin \omega z} (- \frac{\alpha}{\omega} \cos \omega z 
+ \beta)^2,
\label{47}
\end{eqnarray}
which is in general divergent, but only when the equality 
$\beta = -\frac{\alpha}{\omega}$ holds, it becomes
finite. Henceforth, we shall consider this specific case.
Then, it is straightforward to calculate the above integral as well
as the second integral over $z$ in Eq. (\ref{45}) associated with
the KK-mass term. (Note that in $D=5$ there does not exist the third
term in the right-hand side in Eq. (\ref{45}).)  The result is given by
\begin{eqnarray}
S_1^{(0)} &=& -\frac{1}{4} \int d^4 x \sqrt{-\hat{g}} \left[
\frac{\alpha^2}{\omega^3} (-1 + 2 \log 2) \hat{g}^{\mu\nu} 
\hat{g}^{\rho\sigma} f_{\mu\rho}f_{\nu\sigma}
+ \frac{2 \alpha^2}{\omega} \hat{g}^{\mu\nu} a_\mu a_\nu \right]. 
\label{48}
\end{eqnarray}
The quantities in front of the kinetic and the mass terms are obviously 
finite, so the gauge field is localized on an $AdS_4$ brane, which
is to be contrasted with the case of a Minkowski brane \cite{Oda3}.

At this stage, it is worthwhile to examine the mass of the brane gauge
field. Provided that we redefine the brane gauge field $a_\mu$ as
\begin{eqnarray}
\frac{\alpha}{\omega^{\frac{3}{2}}} \sqrt{-1 + 2 \log 2} \ a_\mu 
\rightarrow a_\mu, 
\label{49}
\end{eqnarray}
Eq. (\ref{48}) reads
\begin{eqnarray}
S_1^{(0)} = -\frac{1}{4} \int d^4 x \sqrt{-\hat{g}} \left[
\hat{g}^{\mu\nu} \hat{g}^{\rho\sigma} f_{\mu\rho}f_{\nu\sigma}
+ \frac{2 \omega^2}{-1 + 2 \log 2} \hat{g}^{\mu\nu} a_\mu a_\nu \right].
\label{50}
\end{eqnarray}
{}From this equation, we can read off the mass of the brane gauge field,
which is expressed in terms of the brane cosmological constant 
by using Eqs. (\ref{23}), (\ref{24}) as
\begin{eqnarray}
m^2 = \frac{\omega^2}{-1 + 2 \log 2} = -\frac{1}{3}\frac{1}
{-1 + 2 \log 2} \Lambda_{AdS}.
\label{51}
\end{eqnarray}
The physical condition that the $U(1)$ gauge field $a_\mu$ must be 
$\it{massless}$ 'photon' on an $AdS_4$ brane requires that the brane 
cosmological constant is small enough. It is very intriguing that in 
the present brane model the smallness of the brane cosmological 
constant is directly connected with the smallness of mass of the brane
gauge field, which, we think, is a miracle in the brane world scenario. 
{}From the current experimental data, the bound on the photon mass
is $m < 2 \times 10^{-16} eV$ so our relation (\ref{51}) yields
a weaker constraint on the four dimensional cosmological constant
although it does not conflict with the current experimental data.

As a remark, let us notice that when the equality 
$\beta = -\frac{\alpha}{\omega}$ holds, our solution reduces
to the form
\begin{eqnarray}
u(z)  = - \frac{2 \alpha}{\omega} \cos^2 \frac{\omega z}{2}. 
\label{52}
\end{eqnarray}
Like the graviton considered in the previous section, this solution 
satisfies the Dirichlet condition at $z = \frac{\pi}{\omega}$, 
where $u(z) = 0$. It is quite of interest that the requirement 
of the localization for the gauge field naturally leads to the same 
boundary condition as the other bosonic fields. (We can show that scalar
field also satisfies the same boundary condition 
$z = \frac{\pi}{\omega}$.)

As another remark, let us check if the bulk gauge field is sharply localized
on a brane or spreads rather widely in a bulk. For this, it is useful to 
change the $z$-coordinates to the radial coordinate whose relation can be 
found in Eq. (\ref{27}) and then examine the normalized zero-mode in an
$AdS_4$. In the radial coordinates, the normalized zero-mode in an
$AdS_4$ has the form
\begin{eqnarray}
\hat{u}(r) &=& \frac{1}{\sqrt{I_1^{D=5}}} u(r) \nn\\
&=& -2 \sqrt{\frac{\omega}{-1+2 \log 2}} \frac{1}{1 + \e^{2 \omega r}}. 
\label{53}
\end{eqnarray}
The present observations require $\omega \sqrt{G_N} \ll 1$ where
$G_N$ is the four dimensional Newton's constant, so
it turns out that this zero-mode is not sharply localized but
spreads rather widely in a bulk.

Next, let us consider $D=6, p=4, n=2$ case, that is, a string-like
defect model in six space-time dimensions.
 In this case, we also follow a similar path of arguments 
to the case of $D=5$ domain wall. The concrete expression for $u(z)$
is different between $D = 2k + 3$ and $D = 2k + 4$, so it is valuable
to investigate this simplest case in the branch of $D = 2k + 4$.
 
In the case of $D=6, p=4, n=2$, from
Eq. (\ref{44}) the integral $I_1^{D=6}$ takes the form
\begin{eqnarray}
I_1^{D=6} = \int_{\frac{\pi}{2 \omega}}^{\frac{\pi}{\omega}} dz
\frac{1}{\sin^2 \omega z} (- \frac{\alpha}{4 \omega} \sin 2 \omega z 
+ \frac{1}{2} \alpha z + \beta)^2,
\label{54}
\end{eqnarray}
which is also generally divergent, but only when the equality 
$\beta = -\frac{\alpha \pi}{2 \omega}$ holds, it also becomes
strictly finite. Again, from now on we shall confine ourselves
to this specific case.
Then, after a bit calculations the integral $I_1^{D=6}$ reduces to
\begin{eqnarray}
I_1^{D=6} = \frac{\alpha^2}{16 \omega^3} (1 + 4 \pi \log2 - 
8 \int_0^{\frac{\pi}{2}} d\zeta \zeta \cot \zeta),
\label{55}
\end{eqnarray}
which is finite since the last integral is known to be finite.
Thus, the gauge field is also localized on an $AdS_4$ brane in a
six dimensional space-time with negative cosmological constant.

Evaluating the integrals over $z$, the classical action is of form
\begin{eqnarray}
S_1^{(0)} = -\frac{1}{4} \int d^4 x \sqrt{-\hat{g}} \int d\theta R_0 
\left\{
I_1 \chi^2 \hat{g}^{\mu\nu} \hat{g}^{\rho\sigma} f_{\mu\rho} f_{\nu\sigma}
+ [ \frac{\pi \alpha^2}{2 \omega} \chi^2 + \frac{2 I_1}{R_0^2}
(\partial_\theta \chi)^2 ] \hat{g}^{\mu\nu} a_\mu a_\nu 
\right\}. 
\label{56}
\end{eqnarray}
With respect to integrations over $\theta$, as mentioned before,
they are always finite, whose fact can be shown as follows.
As seen in the derivation from (\ref{41}) to (\ref{42}), $\chi(y^m)$ 
must satisfy the equation of motion, $\partial_m (\sqrt{\tilde{g}} 
\tilde{g}^{mn} \partial_n \chi) = 0$, which now reduces to 
$\partial_\theta^2 \chi = 0$, so a general solution to this equation
is given by $\chi(\theta)= \chi_1 + \theta \chi_2$ where $\chi_1, 
\chi_2$ are integration constants. Thus the integrals over $\theta$
appearing in Eq. (\ref{56}), 
$\int_0^{2 \pi} d\theta \chi^2, \int_0^{2 \pi} d\theta 
(\partial_\theta \chi)^2$ are finite quantities as long as the
integration constants are finite.
We are now ready to examine the mass of the brane gauge field. 
To make the kinetic term take a canonical form, let us redefine the brane 
gauge field $a_\mu$ as follows:
\begin{eqnarray}
\sqrt{R_0 I_1 \int d\theta \chi^2(\theta)} \ a_\mu 
\rightarrow a_\mu. 
\label{57}
\end{eqnarray}
As a result, it turns out that Eq. (\ref{56}) becomes
\begin{eqnarray}
S_1^{(0)} = -\frac{1}{4} \int d^4 x \sqrt{-\hat{g}} \left[
\hat{g}^{\mu\nu} \hat{g}^{\rho\sigma} f_{\mu\rho}f_{\nu\sigma}
+ \frac{8 \pi \omega^2 \int d\theta \chi^2 + \frac{2}{R_0^2} K
\int d\theta (\partial_\theta \chi)^2}{K \int d\theta \chi^2} 
\hat{g}^{\mu\nu} a_\mu a_\nu \right],
\label{58}
\end{eqnarray}
where $K \equiv 1 + 4 \pi \log2 - 8 \int_0^{\frac{\pi}{2}} d\zeta 
\zeta \cot \zeta$. Since it is reasonable to regard value of the 
integrals over $\theta$ as being of the order 1, the smallness of
mass of the brane gauge field requires $\omega^2 \approx 0, 
\frac{1}{R_0^2} \approx 0$, both of which imply that the brane
cosmological constant is extremely tiny as desired. 
 
As a final check, let us study the zero-mode $u(z)$. When the equality 
$\beta = -\frac{\alpha \pi}{2 \omega}$ holds, our solution reduces
to the form
\begin{eqnarray}
u(z)  = - \frac{\alpha}{4 \omega} \sin 2 \omega z + \frac{\alpha}
{2 \omega}(\omega z - \pi). 
\label{59}
\end{eqnarray}
This solution also satisfies the Dirichlet boundary condition at 
$z = \frac{\pi}{\omega}$, where $u(z) = 0$. For the investigation of the
localization of this mode on a brane, we also use the radial coordinates
instead of the $z$-coordinates. In the radial coordinates, after
some calculations, the normalized zero-mode in an $AdS_4$ is proportional to
\begin{eqnarray}
\hat{u}(r) \propto \sqrt{\omega} \Big[\frac{1}{\e^{\omega r} + 
\e^{-\omega r}} \tanh (\omega r) - \tan^{-1}(\e^{-\omega r}) \Big]. 
\label{60}
\end{eqnarray}
Again, for $\omega \sqrt{G_N} \ll 1$, it turns out that the brane gauge 
field is not sharply localized on an $AdS_4$ brane.

\section{Localization of fermionic fields}

Next let us turn to fermionic fields, those are, spin-1/2 spinor field
and spin-3/2 gravitino field. First, let us consider spin-1/2 spinor
field. The starting action is the conventional 
Dirac action with a mass term in D dimensions:
\begin{eqnarray}
S_{1/2} = \int d^D x \sqrt{-g} \bar{\Psi} i (\Gamma^M D_M
+ m \varepsilon(z)) \Psi, 
\label{61}
\end{eqnarray}
where the covariant derivative is defined as $D_M \Psi = ( \partial_M + 
\frac{1}{4} \omega_M^{AB} \gamma_{AB} ) \Psi$ with the definition of
$\gamma_{AB} = \frac{1}{2} [\gamma_A, \gamma_B]$, and $\varepsilon(z)$
is $\varepsilon(z) \equiv \frac{z}{|z|}$ and $\varepsilon(0) \equiv 0$.
Here the indices $A, B$ are the ones of the local Lorentz frame and 
the gamma matrices $\Gamma^M$ and $\gamma^A$ are related by the 
vielbeins $e_A^M$ through the usual relations $\Gamma^M = e_A^M \gamma^A$ 
where  $\{\Gamma^M, \Gamma^N\} = 2 g^{MN}$ and $\{\gamma^A, \gamma^B\} 
= 2 \eta^{AB}$.
A feature of the action is the existence of a mass term with 
a 'kink' profile. We have just introduced this type of mass term in the 
action since the existence has played a critical role 
in the localization of fermionic
fields on a Minkowski brane in an arbitrary dimension \cite{Oda3}.

In this section, we shall consider a more general metric ansatz than 
Eq. (\ref{4}) in the 'radial' coordinates. 
The metric ansatz we take into consideration is the following one:
\begin{eqnarray}
ds^2 &=& g_{MN} dx^M dx^N  \nn\\
&=& e^{-A(r)} \hat{g}_{\mu\nu}(x^\lambda) dx^\mu dx^\nu + dr^2 
+ e^{-B(r)} \tilde{g}_{mn}(y^l) dy^m dy^n,
\label{62}
\end{eqnarray}
where we have replaced a metric on $S^{n-1}$ in Eq. (\ref{4})
with a general curved metric
$\tilde{g}_{mn}(y^l)$ depending only on extra dimensions $y^l$ except $r$.
In this background metric, the torsion-free conditions yield an explicit
expression of the spin connections:
\begin{eqnarray}
\omega_\mu = \frac{1}{4} A'(r) \Gamma_r \Gamma_\mu + 
\hat{\omega}_\mu(\hat{e}), \ \omega_r = 0, \ 
\omega_m = \frac{1}{4} B'(r) \Gamma_r \Gamma_m + 
\tilde{\omega}_m (\tilde{e}),
\label{63}
\end{eqnarray}
where we have defined $\omega_M \equiv \frac{1}{4} \omega_M^{AB} \gamma_{AB}$.
And $\hat{\omega}_\mu(\hat{e})$ and $\tilde{\omega}_m (\tilde{e})$ are
the spin connections constructed out of $\hat{e}_\mu$ and $\tilde{e}_m$,
respectively.
Using Eq. (\ref{63}), the Dirac equation $(\Gamma^M D_M + m \varepsilon(r)) 
\Psi = 0$ can be cast to the form
\begin{eqnarray}
\left[\Gamma^r (\partial_r - \frac{p}{4} A' - \frac{n-1}{4} B')  
+ \Gamma^\mu (\partial_\mu + \hat{\omega}_\mu) 
+ \Gamma^m (\partial_m + \tilde{\omega}_m) + m \varepsilon(r) 
\right] \Psi = 0. 
\label{64}
\end{eqnarray}
Let us find the massless zero-mode solution with the form of $\Psi(x^M) = 
\psi(x^\mu) u(r) \chi(y^m)$ such that 
$\Gamma^\mu \hat{D}_\mu \psi = \Gamma^m \tilde{D}_m \chi = 0$ and the 
chirality condition $\Gamma^r \psi = \psi$ is imposed on the brane fermion. 
Then, Eq. (\ref{64}) is reduced to a first-order differential equation 
for $u(r)$ and is easily solved to be 
\begin{eqnarray}
u(r) = u_0 \e^{\frac{p}{4} A(r) + \frac{n-1}{4} B(r) - m \varepsilon(r) r},
\label{65}
\end{eqnarray}
with an integration constant $u_0$. 

In order to check the localization of this mode, let us
plug this solution into the Dirac action (\ref{61}). Then the action 
reduces to the form
\begin{eqnarray}
S_{1/2}^{(0)} &=& \int d^D x \sqrt{-g} \bar{\Psi}^{(0)} i (\Gamma^M D_M
+ m \varepsilon(z)) \Psi^{(0)} \nn\\
&=& u_0^2 \int d^{n-1} y \sqrt{\tilde{g}} \chi^\dagger(y) \chi(y)
\int_0^{\infty} dr e^{\frac{1}{2}A(r) - 2m \varepsilon(r) r} 
\int d^p x \sqrt{-\hat{g}} \bar{\psi} i \gamma^\mu \hat{D}_\mu \psi
+ \cdots. 
\label{66}
\end{eqnarray}
The condition of the trapping of the bulk spinor on an $AdS_p$ brane 
requires that an integral over $r$ has a finite value since an integral
over $y$ is finite. The integral is easily evaluated as follows:
\begin{eqnarray}
I_{1/2} &=& \int_0^{\infty} dr e^{\frac{1}{2}A(r)- 2m \varepsilon
(r) r}  \nn\\
&=& \int_0^{\infty} dr \frac{1}{\cosh \omega r} e^{- 2m \varepsilon
(r) r}.
\label{67}
\end{eqnarray}
This integral is obviously finite so the bulk spinor is confined
to a brane by the gravitational interaction. In particular, in the 
case of massless fermion, the above integral can be integrated to be
\begin{eqnarray}
I_{1/2}^{m=0} = \frac{\pi}{2 \omega}.
\label{68}
\end{eqnarray}
Namely, even in the massless spinor, the bulk spinor is localized
on a brane, whose fact should be contrasted with the case of a Minkowski
brane, where only the massive bulk fermion with a 'kink' profile is 
localized on the brane whereas the massless one is not so. 
This fact can be traced in Eq. (\ref{68}) since $I_{1/2}^{m=0}$ at 
$\omega = 0$ is divergent.
(Note that in both the Minkowski brane and the AdS brane, the form of 
the zero-mode solution of fermion, (\ref{65}), is the same so this 
consideration is legitimate.) In the case at hand, irrespective
of the presence of mass term, the bulk spinor can be localized on
the brane through the gravitational interaction as long as the
brane cosmological constant is nonvanishing.

However, there is a caveat. As seen in the concrete form of
$u(r)$ in (\ref{65}), the zero-mode $u(r)$ can be written to
\begin{eqnarray}
u(r) &=& \frac{u_0}{R_0^{\frac{n-1}{2}}} 
(\cosh \omega r)^{-\frac{D-1}{2}} \e^{- m \varepsilon(r) r} \nn\\
&\approx& \frac{u_0}{R_0^{\frac{n-1}{2}}} 2^{\frac{D-1}{2}}
\e^{- \frac{D-1}{2} \omega r - m \varepsilon(r) r}.
\label{65A}
\end{eqnarray}
The last expression was derived under the condition $\omega r \gg 1$.
This expression tells us that provided that $\omega \sqrt{G_N} \ll 1$, 
the zero-mode spreads more widely in a bulk in the massless
limit. To avoid such a situation, we might also need the presence
of a mass term with 'kink' profile.
Incidentally, note that the results obtained so far are independent of
a concrete form of $B(r)$, so only the warp factor $\e^{-A(r)}$
in front of the $p$-dimensional metric controls the results.

Next, let us consider the gravitino field of spin-3/2. 
The action for the spin-3/2 bulk gravitino is given by the Rarita-Schwinger 
action
\begin{eqnarray}
S_{3/2} = \int d^D x \sqrt{-g} \bar{\Psi}_M i \Gamma^{[M} \Gamma^N
\Gamma^{R]} (D_N + \delta_N^r \Gamma_r m \varepsilon(r)) \Psi_R, 
\label{69}
\end{eqnarray}
where $D_M \Psi_N = \partial_M \Psi_N - \Gamma^R_{MN} \Psi_R + 
\frac{1}{4} \omega_M^{AB} \gamma_{AB} \Psi_N$ and the square bracket
denotes the anti-symmetrization with weight 1. From the metric
condition $D_M e_N^A = \partial_M e_N^A - \Gamma^R_{MN} e_R^A + 
\omega_M^{AB} e_{NB} = 0$, we obtain the concrete expression for the
affine connections  $\Gamma^R_{MN} = e^R_A (\partial_M e_N^A 
+ \omega_M^{AB} e_{NB})$. 
With the gauge condition $\Psi_r = 0$ and assuming $\Psi_m = 0$
for simplicity, 
the equations of motion $\Gamma^{[M} \Gamma^N \Gamma^{R]} (D_N + 
\delta_N^r \Gamma_r m \varepsilon(r)) \Psi_R = 0$ can be cast to 
the form 
\begin{eqnarray}
g^{\mu\nu} \left[\Gamma^r (\partial_r - \frac{p-2}{4} A'(r) 
- \frac{n-1}{4} B'(r)) + m \varepsilon(r)
\right] \Psi_\nu = 0, 
\label{70}
\end{eqnarray}
where we have used equations $\gamma^\mu \Psi_\mu = \hat{D}^\mu \Psi_\mu
= \gamma^{[\mu} \gamma^\nu \gamma^{\rho]} \hat{D}_\nu \Psi_\rho = 
\Gamma^m \tilde{D}_m \Psi_\mu = 0$.
Let us look for a solution with the form $\Psi_\mu(x^M) = \psi_\mu 
(x^\lambda) u(r) \chi(y^m)$. If the chirality condition $\Gamma^r \psi_\mu = 
\psi_\mu$ is utilized in Eq. (\ref{70}), we can get a solution
\begin{eqnarray}
u(r) = u_0 \e^{\frac{p-2}{4} A(r) + \frac{n-1}{4} B(r) - m \varepsilon(r) 
r},
\label{71}
\end{eqnarray}
with an integration constant $u_0$. 

Substituting this solution into the action (\ref{69}), we arrive at 
the following expression
\begin{eqnarray}
S_{3/2}^{(0)} &=& \int d^D x \sqrt{-g} \bar{\Psi}_M^{(0)}
i \Gamma^{[M} \Gamma^N \Gamma^{R]} (D_N + \delta_N^r \Gamma_r m 
\varepsilon(r)) \Psi_R^{(0)} \nn\\
&=& u_0^2 \int d^{n-1} y \sqrt{\tilde{g}} \chi^2(y) 
\int_0^{\infty} dr e^{\frac{1}{2} A(r) - 2m \varepsilon(r) r} 
\int d^p x \sqrt{-\hat{g}} \bar{\psi}_\mu i \gamma^{[\mu} 
\gamma^\nu \gamma^{\rho]} \hat{D}_\nu \psi_\rho
 + \cdots. 
\label{72}
\end{eqnarray}
Again the condition for the localization of the gravitino on a
brane requires the integral over $r$ to take a finite value.
Namely, the following integral over $r$ must be finite 
\begin{eqnarray}
I_{3/2} &=& \int_0^{\infty} dr e^{\frac{1}{2}A(r)- 2m \varepsilon
(r) r}  \nn\\
&=& \int_0^{\infty} dr \frac{1}{\cosh \omega r} e^{- 2m \varepsilon
(r) r}.
\label{73}
\end{eqnarray}
Here let us notice that this condition has the same form as in spin 1/2 
spinor field, Eq. (\ref{67}), so the spin 3/2 gravitino is also localized 
on a brane. The form of $u(r)$ in (\ref{71}), however, is similar to
that of spin-1/2 spinor field (\ref{65}), so as in the spinor field
it might be necessary to include a mass term with a 'kink' profile
in order to have a sharp localization on a brane.

\section{Discussions}

In this article we have discussed locally localized gravity models
in higher dimensions. As a solution to Einstein's equations with
a set of scalar fields with $\it{global}$ $SO(n)$ symmetry, we
have found two types of $AdS_p$ brane solution in a unified metric
ansatz. Though these solutions have been already found in 
Ref. \cite{Vilenkin}, our derivation is more concise than
their derivation and we furthermore spelled out the physical properties
of the solutions. An important issue that we have found in this paper
is that in higher dimensions the solutions which are free from the naked
curvature singularity and possess the property of gravity localization
are very few. Apart from a type of trivial extension of the Randall-Sundrum
solutions, in higher dimensions the physical solution corresponds to only 
an $AdS_p$ brane in a space-time with negative bulk cosmological constant. 
It is quite curious that there are no nontrivial solutions in higher 
dimensions which correspond to a $dS_p$ brane and a $M_p$ brane solution 
with needed physical properties.
From this point of view, more study about an $AdS$ brane seems to be
deserved in future. In higher dimensions, intersection brane solutions 
with two different warp factors might be needed in order to satisfy the
physical properties \cite{Sakai}.

Concerning the localization of various bulk fields on an $AdS$
brane only by the gravitational interation, we have explicitly considered 
spin-1/2 spinor, spin-1 vector, and spin-3/2 gravitino fields. We have also
implicitly considered spin-2 graviton where we have stressed that a complete 
understanding of the gravity localization requires us to find a reasonable
core model. The local fields which we have left aside are spin-0 scalar and
higher-rank antisymmetric tensor fields. It is well known that a real scalar
field satisfies the same equation of motion as that of the transverse,
traceless graviton modes, so a real scalar field shares the common
localization
properties with the graviton. The treatment of higher rank tensor fields is
completely parallel to that of the gauge fields, so we have skipped these
cases.

It is worth stressing here that the localization mechanism, that we have 
found in this paper, in particular,
is new and novel for spin-1 vector and fermionic fields. For the former,
it is known that the gauge field is not localized on a Minkowski brane
in the original Randall-Sundrum model. On the other hand, in an anti-de
Sitter 
brane, the gauge field $\it{is}$ localized due to the presence of the brane
cosmological constant. However, there is a caveat. Namely, although the gauge
field is anyway localized near the brane, it is not sharply localized, by
which we meet some phenomenological problems such as the violation of the 
charge conservation law in our world.
Moreover, we have found a new phenomenon that the size 
of the brane cosmological constant is determined by that of mass of 
'$\it{photon}$' on a brane. Also for the fermionic fields, the presence of
the brane cosmological constant provides a novel localization mechanism
where massless fermions are localized on an $AdS$ brane whereas
only massive fermion with a 'kink' profile can be localized on a $M$ brane
as in the Randall-Sundrum model. However, as in the gauge field, 
the zero-mode of massless fermions also spreads
rather widely in the bulk space-time, so we might need a mass term with
a 'kink' profile to have a sharply localized brane fermion.

Of course, if we wish to construct a fully successful brane world model
in higher dimensions on the basis of global defects, it is essential to
understand physics inside the core of the defects. Without knowledge of it, 
we cannot fully answer several questions such as stability of the defects.
Another unsolved problem within the context of the present formulation
is how to construct a model with two or more branes, which would be
necessary to understand the mass hierarchy problem between the Planck
scale and the electro-weak scale. 
There are also future works of the construction of a supersymmetric
model corresponding to the present model and of deriving the model
at hand from superstring theory. We wish to clarify these important
problems in a future publication.  

\vs 1


\end{document}